\documentclass[aps,preprint,prX,nofootinbib,floatfix,showpacs,preprintnumbers]{revtex4} 
\usepackage{graphicx}
\input epsf
\newcommand{\sfig}[2]{
\includegraphics[width=#2]{#1}
        }
\newcommand{\Sfig}[2]{
    \begin{figure}[thbp]
    \sfig{#1.eps}{\columnwidth}
    \caption{{\small #2}}
    \end{figure}
}

\newcommand{\rf}[1]{\ref{fig:#1}}

\def\lsim{\mathrel{\raise.3ex\hbox{$<$\kern-.75em\lower1ex\hbox{$\sim$}}}}
\def\gsim{\mathrel{\raise.3ex\hbox{$>$\kern-.75em\lower1ex\hbox{$\sim$}}}}

\def\cmm2{{\,\rm cm^{-2}}}
\def\cm2{{\,{\rm cm}^2}}
\def\cmm3{{\,{\rm cm}^{-3}}}
\def\gcmm3{{\,{\rm g\,cm^{-3}}}}

\def\fun#1#2{\lower3.6pt\vbox{\baselineskip0pt\lineskip.9pt
  \ialign{$\mathsurround=0pt#1\hfil##\hfil$\crcr#2\crcr\sim\crcr}}}

\def\be{\begin{equation}}
\def\ee{\end{equation}}
\def\bea{\begin{eqnarray}}
\def\eea{\end{eqnarray}}
\newcommand{\vs}{\nonumber\\}

\newcommand{\ec}[1]{Eq.~(\ref{eq:#1})}
\newcommand{\Ec}[1]{(\ref{eq:#1})}
\newcommand{\eql}[1]{\label{eq:#1}}

\begin{document}

\preprint{FERMILAB-PUB-09-063-A, CERN-PH-TH/2009-030}

\title{Constraining Cosmological Dark Matter Annihilation with Gamma Ray Observations}

\author{Scott Dodelson$^{1,2,3}$, Alexander V.~Belikov$^{4}$, Dan Hooper$^{1,2}$, and Pasquale Serpico$^{1,5}$}

\affiliation{$^1$Center for Particle Astrophysics, Fermi National
Accelerator Laboratory, Batavia, IL~~60510-0500, USA}
\affiliation{$^2$Department of Astronomy \& Astrophysics, The
University of Chicago, Chicago, IL~~60637-1433, USA}
\affiliation{$^3$Kavli Institute for Cosmological Physics, Chicago, IL~~60637-1433, USA}
\affiliation{$^4$Department of Physics, The
University of Chicago, Chicago, IL~~60637-1433, USA}
\affiliation{$^5$Physics Department, Theory Division, CERN, CH-1211 Geneva 23, Switzerland}
\date{\today}
\begin{abstract}
Annihilation of cosmologically distributed dark matter is predicted to produce a potentially observable flux of high energy photons. 
This signal is predicted to be virtually uniform on the sky but, in order to be identified, 
must be extracted from various Galactic and extragalactic backgrounds. We consider three techniques for extracting this signal from the backgrounds: spectral discrimination,
angular discrimination, and {\it distribution} discrimination. We analyze the first two of these with the Fisher Matrix formalism to obtain projections
for constraints from the Fermi satellite. The third technique exploits the fact that the number of 
photons from extragalactic blazars is drawn from a distribution which is far from Poisson. 
Using a toy model, we show that knowledge of this distribution enhances one's ability to extract the dark matter signal, while ignorance of it 
can lead to the introduction of a large systematic error. 
\end{abstract}
\maketitle

\section{Introduction}

There is abundant evidence that non-baryonic dark matter is responsible for many gravitational effects
observed over a wide range of scales~\cite{Bertone:2004pz}. Experimental efforts are now focused on identifying the particle nature of this substance. A particularly interesting possibility is that the dark matter may take the form of a weakly interacting massive particle (WIMP) which could be observed in underground direct detection experiments~\cite{Aprile:2005ww,Sanglard:2006hd,Bisset:2007mz,Bruch:2007zz} and/or be produced at accelerators such
as the Large Hadron Collider~\cite{Baltz:2006fm}. A third class of experimental approaches 
to this problem, known as indirect detection, consists of experiments which search 
for the products of dark matter annihilations, including neutrinos, cosmic rays, and gamma rays. 

A new and exciting range of possibilities for the indirect detection of dark matter 
has been opened with the launch of the satellite-based Fermi gamma ray space 
telescope (formerly known as GLAST)~\cite{Bergstrom:1997fj,Cecchi:2008zz}. 
Fermi is sensitive to photons in the 100 MeV-300 GeV range, and benefits 
from far greater exposure and superior angular and energy resolution than 
its predecessor, EGRET.  The flux of gamma rays produced in dark matter 
annihilations depends on both the WIMP's annihilation cross section, 
mass, and dominant annihilation modes, and on the spatial distribution 
of dark matter. An advantage of indirect detection relative to direct 
detection efforts is that the annihilation cross section probed is 
in many models directly related to that responsible for the primordial 
abundance of dark matter. Although there is variation from model-to-model, annihilation cross sections of order $\langle\sigma v\rangle \sim 3\times 10^{-26}$ cm$^{3}$ sec$^{-1}$ 
are common across a wide range of dark matter candidates. If the dark matter annihilation 
cross section is of this magnitude, Fermi and
ground-based gamma ray telescopes will likely detect many 
photons from dark matter. The challenge lies in separating this signal from astrophysical backgrounds, 
which are likely to be tens to thousands of times as large, depending on the energy bin and direction on the sky. 

A general strategy for optimizing the chances of detecting dark matter is to combine angular and spectral features to disentangle the signal from backgrounds. The details of how this is best done, however, depend on the specific target one is focusing on. 
For example, in previous work~\cite{Dodelson:2007gd}, three of us discussed techniques for separating dark matter annihilation products from astrophysical backgrounds in the Galactic Center region. 
The angular features of the signal from the smooth Galactic halo, or from unresolved sub-halos, may also provide useful information for signal/background discrimination, either in real or multipole space~\cite{Miniati:2007ke,Hooper:2007be,SiegalGaskins:2008ge,Fornasa:2009qh,SiegalGaskins:2009ux}. 

A different situation holds for the diffuse gamma ray flux resulting from the integrated sum of all extragalactic dark matter halos (the {\it cosmological} signal). 
To be identified,  this signal will have to be separated from the extragalactic background 
due to unresolved gamma ray sources, such as blazars, as well as from residual contamination from the Galaxy.  This procedure is delicate and, 
not surprisingly, the astrophysical interpretation of the results in the case of EGRET data has led 
to very different conclusions, see e.g.~\cite{Sreekumar:1997un,Strong:2004ry,Elsaesser:2004ap,deBoer:2006tv}. Also, 
when removing the ``Galactic background'' one must account for the DM signal:  Under some common assumptions
(universality of the DM profile in the halos) this signal is expected to dominate
over the extragalactic one~\cite{Ando:2005hr,Hooper:2007be}. Still, the cosmological DM signal is subject
to very different systematics compared to the Galactic one and encodes a lot of information on the cosmological properties of DM, 
justifying a deeper study. Apart from the angular distribution of both signal and 
background ~\cite{Ando:2005xg,Ando:2006mt,Cuoco:2006tr,Ando:2006cr,Cuoco:2007sh,Taoso:2008qz}, there remain two potential 
differences which can be exploited to extract the signal:
\begin{itemize}
\item The energy spectra of the signal and background are likely to be quite different. This difference has often been exploited to determine how well the signal can be extracted. In this paper, we use the Fisher Matrix formalism to simplify this task.
\item{A common assumption underlying previous work has been that the number of photons from both signal and background in a given angular pixel are drawn from a Poisson distribution. In fact, as we illustrate in~\S{II}, this is not true in general. In particular,  the blazar-produced photons are likely to be drawn from a probability distribution function (PDF) very different than Poisson.  This opens the possibility of using the different underlying distributions to separate signal from background. Recently, a similar statistic has been studied for use in characterizing the signal of unresolved Galactic dark matter sub-halos~\cite{Lee:2008fm}.}
\end{itemize}

In this paper, we explore the efficiency of these techniques applied to pixel-statistics for extracting the gamma ray flux from cosmological 
dark matter annihilations. We derive a compact way to assess how effectively a given experiment can separate signal from background using 
spectral information alone (\S{III}) and then using both spectral and angular information (\S\ref{sec:ani}). In \S{V}, we explore
the information encoded in yet another potential discriminant: the probability distribution function (PDF) of counts.
We make a simple 
attempt to understand the different distributions and find that there are both large advantages if one uses the correct 
distribution and considerable disadvantages if one assumes an incorrect distribution (\S{IV}). A discussion and our conclusions are reported in~\S{V}. 

\section{Models of the Signal and Background}
\label{sec:model}
Here, we describe simple models for the 
dark matter annihilation signal, for the background from unresolved blazars, and the Galactic background.

\subsection{Cosmological Dark Matter Signal}

It has long been realized that, due to the clumpiness of virialized dark matter structures, the extragalactic dark matter annihilation signal is much larger than its naive expectation value from the average dark matter abundance in the universe~\cite{Silk:1992bh}. The flux of gamma rays produced in dark matter annihilations throughout the cosmological volume is described by
\begin{eqnarray}
\frac{d\phi_{\gamma}}{dE_{\gamma, 0}}&=& \frac{\langle \sigma v \rangle}{8 \pi}\frac{c}{H_0}\frac{\bar{\rho}^2_{X}}{m^2_X} \int dz (1+z)^3 \frac{\Delta^2(z)}{h(z)}  \nonumber \\
&&\times \frac{dN_{\gamma}}{dE_{\gamma}}(E_{\gamma}(1+z)) e^{-\tau(z, E_{\gamma})},\label{eq:phi}
\end{eqnarray}
where $\langle \sigma v \rangle$ and $m_X$ are the annihilation cross section and mass of the WIMP. 
The spectrum of gamma rays per annihilation, $dN_{\gamma}/dE_{\gamma}$, further depends on the dominant annihilation channels.  
In this study, we consider the case of a 100 GeV WIMP which annihilates uniquely to $W^+ W^-$ with cross section 
$\langle\sigma v\rangle=3\times 10^{-26}$ cm$^3$\,sec$^{-1}$, which in turn produce gamma 
rays through their decays. In Eq.~(\ref{eq:phi}), $\bar{\rho}_X$ denotes the average density of 
dark matter, $\Delta^2(z)$ the average squared overdensity, $\tau$ describes the estimated optical depth of the universe to 
gamma rays, $H_0=70$ km/s/Mpc is the present value of the Hubble constant and $h(z)\equiv \sqrt{(1+z)^3\Omega_M+\Omega_\Lambda}$ describes 
its evolution with redshift $z$ in terms of the matter fraction, $\Omega_M=0.3$, and cosmological constant, $\Omega_\Lambda=1-\Omega_M$ 
(a flat universe is assumed). To calculate the flux of gamma rays from WIMP annihilations, we follow the procedure of Ref.~\cite{Ullio:2002pj}, 
assuming  a universal halo profile either of the Navarro, Frenk and White (NFW)~\cite{Navarro:1996gj} or Moore {\it et al.}~\cite{Moore:1999nt} 
form.  We adopt the Bullock {\it et al.}~\cite{Bullock:1999he} convention for estimating halo concentrations, which leads to  
enhancement factors of $\Delta^2(0)=1.15\times 10^5$ and $1.18\times 10^6$ for the two
models, respectively. 

An important caveat is in order:
Clearly, towards the Galactic Center this is {\it not} the dominant component of the diffuse dark matter signal, 
since the signal from the smooth halo of our Galaxy is larger. At high Galactic latitudes (which constitute 
the largest fraction of the solid angle), the signal which dominates depends on the degree of substructure 
surviving in the Milky Way~\cite{Hooper:2007be}. Calculations based on recent simulations~\cite{Fornasa:2009qh} 
suggest that the dark matter signal from galactic substructure dominate the (quasi-)isotropic background, 
at least for typical substructure distributions inferred from pure dark matter N-body simulations. 
Yet, quite a bit of uncertainty remains, especially since baryonic effects have not yet been included.
Here, for simplicity, we consider only the extragalactic component, keeping in mind that for
a given choice of the halo profile, this may {\it underestimate} the real contribution to the signal.

\subsection{Unresolved Blazars}

Over its mission, the EGRET experiment accumulated a catalog of 66 blazars (at high confidence)~\cite{Hartman:1999fc,Mukherjee:1997qw}. 
From the information contained in this catalog, it is possible to construct a model of the redshift distribution, luminosity function, 
and spectrum of these sources. In turn, such a model can be used to estimate the total flux of gamma rays expected to be produced by the 
large population of unresolved (typically fainter, or more distant) blazars. In this analysis, we adopt a blazar luminosity function based 
on the population study of Ref.~\cite{luminosityfunction}, and use a redshift distribution following the sub-mm/far-IR luminosity density 
associated with luminous IR galaxies~\cite{Dermer:2006pd}. We also adopt a universal spectral shape of $dN_{\gamma}/dE_{\gamma} \propto E_{\gamma}^{-2.2}$. 

Although this model is broadly consistent with the properties of the blazars observed by EGRET, 
the limited sample size present in the EGRET catalog (and the limited amount of information available for each blazar) 
makes it difficult to construct such a model with much accuracy. This situation will be dramatically improved as Fermi 
begins to accumulate its own catalog of blazars. In particular, Fermi is expected to resolve $\sim10^3$ blazars, 
providing a much larger sample with which to perform population studies.  In fact, 104 blazars have already been 
detected with very high confidence ($\gsim10 \sigma$) in the first 90 days of Fermi data~\cite{Abdo:2009wu}. 
Furthermore, these observations will extend to much higher energies than those of EGRET, and will include 
blazars with lower luminosities and higher redshifts. These observations will enable the construction of a
 population model which will be capable of estimating the diffuse gamma ray spectrum from (unresolved) blazars with far greater accuracy than is currently possible. 

In Fig.~\ref{fig:spectra}, we compare the diffuse gamma ray spectrum from unresolved blazars in our model with 
that from dark matter annihilations with the parameters assumed above. The flux from dark matter is shown 
for the case of both NFW and Moore {\it et al.} profiles. Note that only the normalization and not the spectral 
shape is affected by the choice of halo profile. Shallower dark matter halo profiles or a decrease in small-scale 
substructure would lower the signal, while any residual contribution from unresolved substructure at high galactic 
latitudes would boost it. A similar enhancement could result due to a larger cross section or additional small scale structures. 

\Sfig{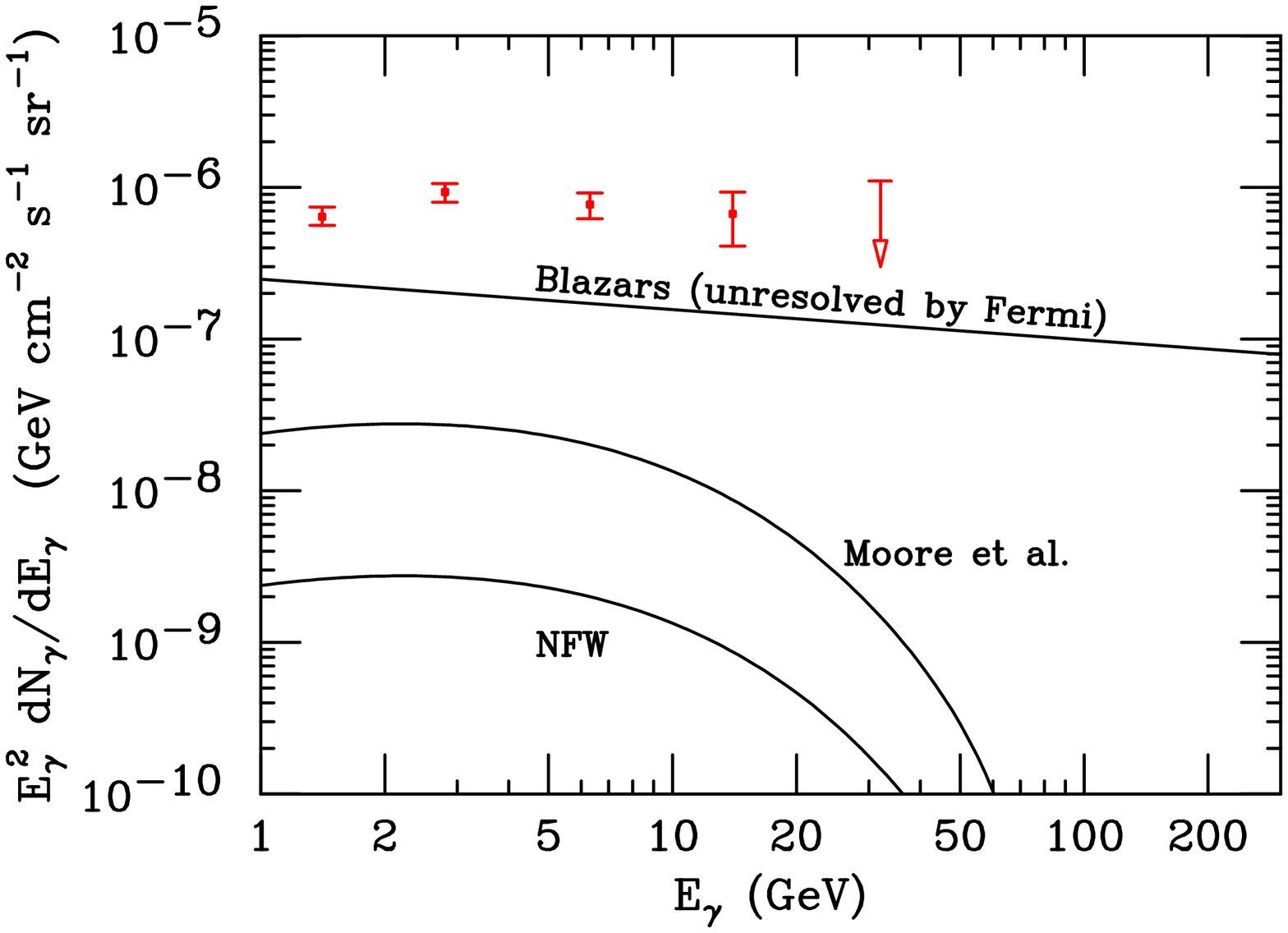}{The cosmological diffuse spectrum of gamma rays from dark
matter annihilations and from unresolved blazars (from Ref.~\cite{luminosityfunction} which may have suffered from incompleteness). 
We have considered a WIMP with a mass of 100 GeV, an annihilation cross section 
of $\langle \sigma v \rangle = 3 \times 10^{-26}$ cm$^3$\,sec$^{-1}$, and which annihilates to $W^+ W^-$. Results are shown for 
two choices of the halo profile (NFW~\cite{Navarro:1996gj} and Moore {\it et al}~\cite{Moore:1999nt}). For details regarding 
our blazar model, see the text. Also shown for comparison is the extragalactic diffuse flux observed by EGRET, as calculated 
in Ref.~\cite{Strong:2004ry}, and an estimate of its fraction that will not be resolved by Fermi.\label{fig:spectra}}

Eq.~(\ref{eq:phi}) represents the average flux on the sky from cosmological dark matter annihilations. For any given 
experiment, this can be turned into the expected numbers of photons per pixel over a finite time. For example, imagine 
dividing half of the sky (the half least contaminated by the Galaxy) into $N_{\rm pix}=330,000$ spatial pixels, each 
roughly ($0.25^\circ)^2$, and counting the number of photons in each pixel accumulated over 5 years of observations 
with the Fermi satellite. Under the assumptions laid out above, Fermi would detect on average $0.06$ photons per pixel 
(over 19,000 total photons over half of the sky) from cosmological dark matter annihilations, assuming an NFW profile. 
The mean count per pixel, in this case 0.06, does not tell the whole story, however. There is also the distribution 
from which photon counts in each pixel are drawn. Strictly speaking, neither the dark matter signal nor the blazar 
background are drawn from a truly Poisson distribution. Yet, the dark matter distribution is much more similar to 
Poisson, because there are many dark matter halos, most of which produce only one or no detectable photons over 
the duration of the experiment. Most halos generate zero photons, some produce one, few produce two, etc.  

The photon counts from blazars are drawn from a very different distribution, however, because only a 
small fraction of halos (those with aligned Active Galactic Nuclei) host blazars. Compared to dark 
matter halos, a larger fraction of these blazars are expected to produce many photons. 
Using information from the EGRET satellite, we can construct a model of blazar-produced photons and 
compare the distribution from which these are drawn to a Poisson distribution. Note that here 
we are making two (probably unrealistic) approximations: (i) We are considering the case where 
the only background is due to blazars. While it is likely that emission from blazars makes up a 
large fraction of the isotropic flux, obviously this is a simplification. (ii) We are 
considering the dark matter signal as Poisson-distributed, which might be valid only 
for a fraction of the signal. Still, in order to illustrate the point, it is useful to
work with these assumptions. In \S{V} we shall come back discussing qualitatively the impact of relaxing these approximations.

\Sfig{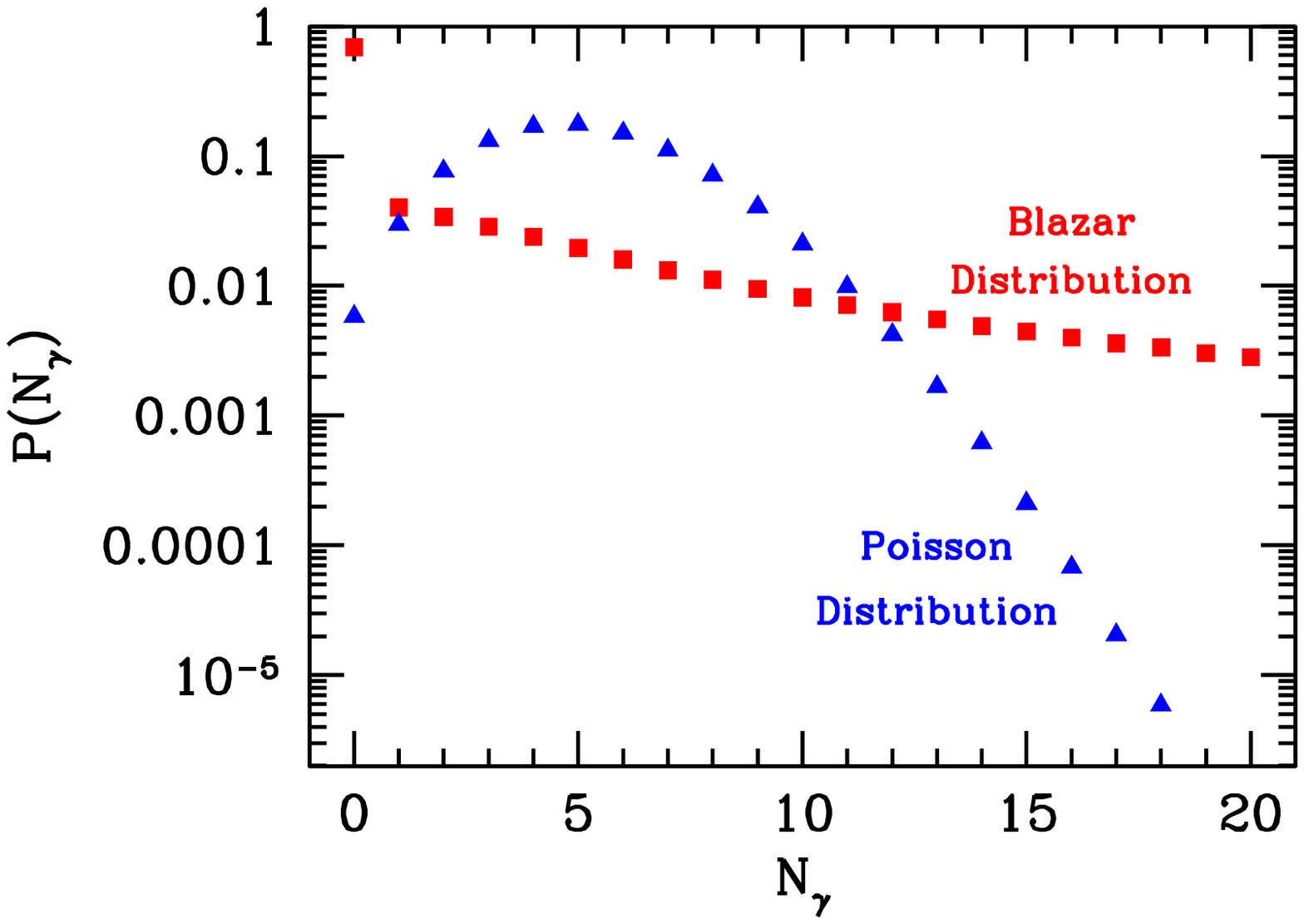}{The probability of observing $N_\gamma$ photons above 1 GeV in a $(0.25^{\circ})^2$ pixel in 5 years of Fermi observations. 
The Poisson distribution is normalized to give the same number of total photons. Note the large tail in blazar distribution compared with a 
Poisson distribution.\label{distr}}

In Fig.~\ref{distr}, we show the probability distribution for unresolved blazars in our model to produce $N_\gamma$ 
detected photons in a given angular pixel of Fermi over 5 years. This is compared with a Poisson distribution which 
has the same number of expected photons, $\sum_{N_\gamma} N_\gamma
P(N_\gamma)$. The key point is that these two distributions are very different from one another; in particular, the 
blazar distribution leads to many more pixels with many photons relative to the corresponding Poisson distribution. 
The total number of photons due to unresolved blazars in this model is $1.7 \times 10^6$, nearly 100 times the number 
produced by dark matter annihilations using an NFW profile.

\begin{figure}[thbp]
\vspace{-40pt}
    \sfig{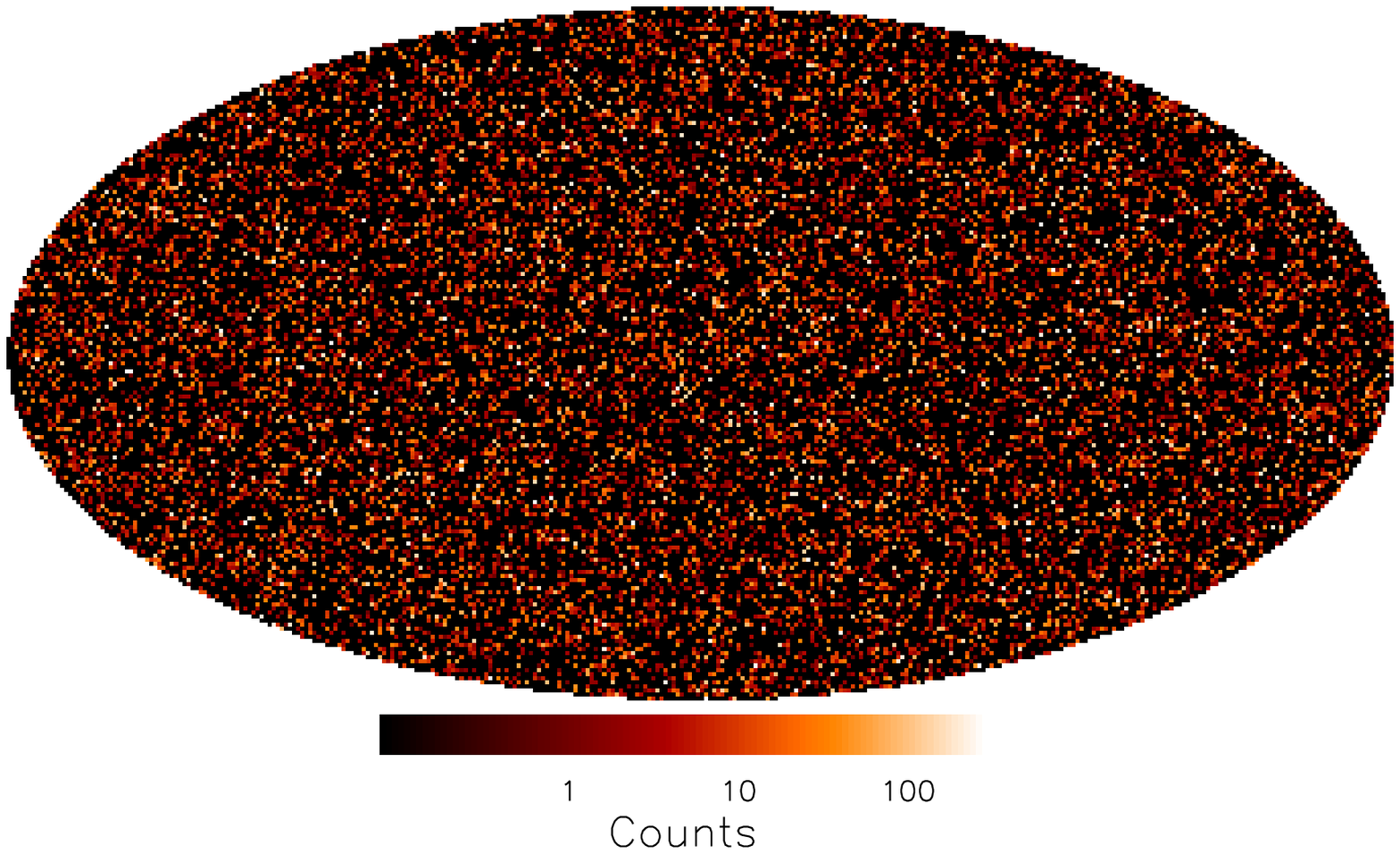}{\columnwidth}
\vspace{-10pt}
\sfig{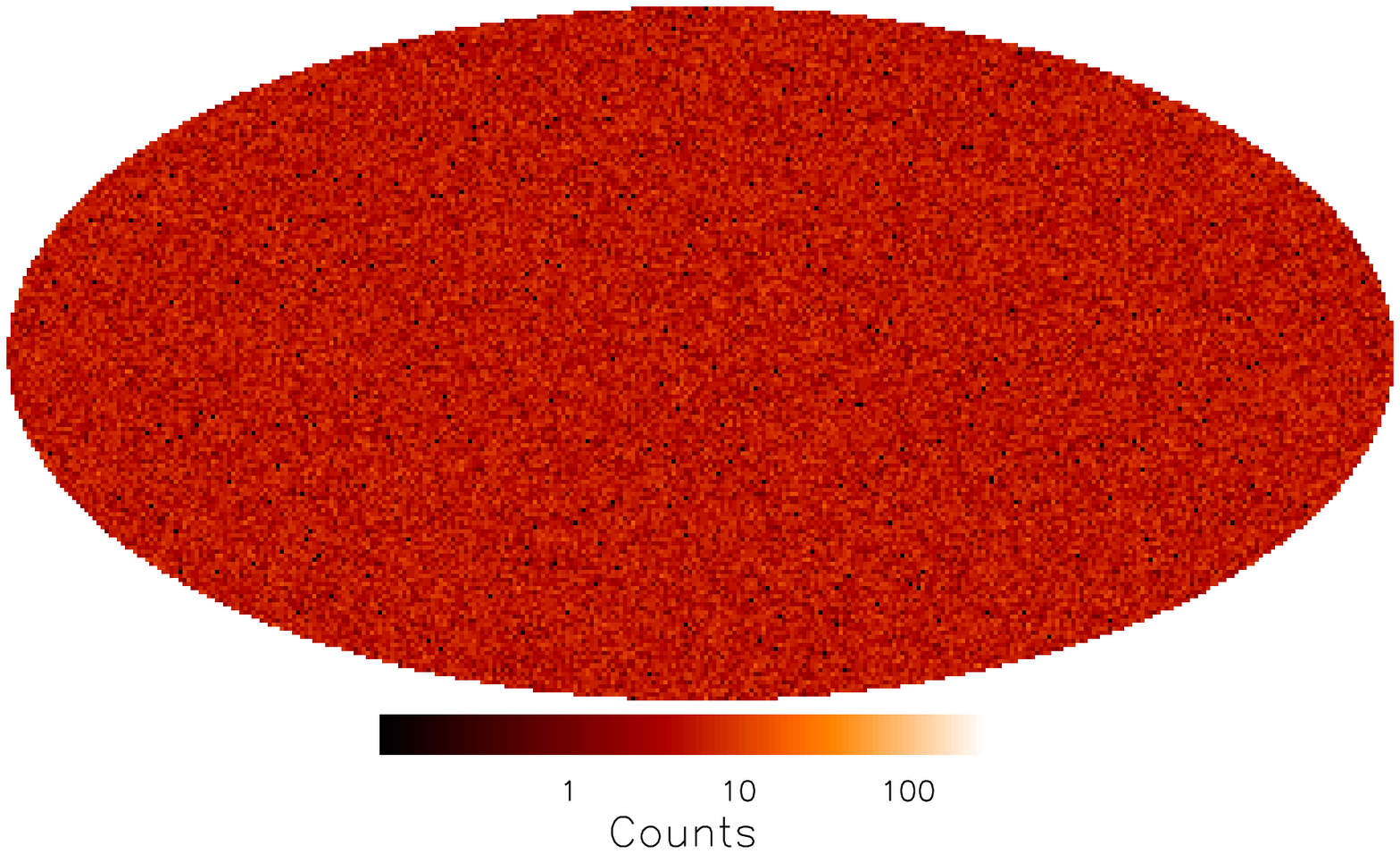}{\columnwidth}
    \caption{{\small Top: Map of counts from unresolved blazars using blazar model described in the text. Bottom: Map of the same number of total counts drawn from a Poisson distribution. }}
    \label{map}
    \end{figure}

In Fig.~\ref{map}, we depict these distributions in two maps containing photons only from unresolved blazars. 
The photons in each pixel in the top map are drawn from the model distribution depicted in Fig.~\ref{distr}. 
There are many pixels with no photons (no blazars in that direction), but some pixels contain several hundred 
photons (pixels with more than 220 photons are considered to be resolved and hence eliminated from the map). 
In contrast, in the bottom frame we show the map corresponding to photons drawn from a Poisson distribution 
with the same number of photons per pixel as in the top map. The multiplicity in the Poisson distribution 
map is much more even: relatively few pixels with either no photons or with $N_\gamma>10$. 
This provides us with a new tool for discriminating the dark matter signal from background: the PDF of observed photons. 

\subsection{Galactic Background}

Even far from the Galactic plane, the Galactic background is considerably larger than the dark matter signal so must be included to obtain realistic 
projections. 
A simple fit, proposed in \cite{Bergstrom:1997fj} and calibrated on EGRET data, for the intensity of photons from the Galaxy as a function
of energy and Galactic coordinates 
is~\cite{Hunter:1997}
\begin{equation}
I_{\rm gal}(E,l,b)=N_0(l,b)\,I_0(E) 
 \label{spectrumGal}
\end{equation}
where  
\begin{equation}
I_0(E)\equiv 10^{-6}
\left(\frac{E}{{\rm GeV}}\right) ^{-2.7}\,{\rm cm}^{-2}{\rm
s}^{-1}{\rm sr}^{-1}{\rm GeV}^{-1} ,\end{equation}
and
\begin{equation}\label{linsys1}
N_{0}(l,b)\equiv\left\{
\begin{array}{cc}
\frac{85.5}{\sqrt{1+(l/35)^2}\sqrt{1+[b/(1.1+0.022\,|l|)]^2}}+0.5&\:\:|l|\geq 30^{\circ}\\
\frac{85.5}{\sqrt{1+(l/35)^2}\sqrt{1+(b/1.8)^2}}+0.5&\:\:|l|\leq
30^{\circ}
\end{array}
\right.
\end{equation}
and both $l$ and $b$ are in degrees.

This model predicts that Fermi will detect $6.1\times 10^7$ photons above 1 GeV from the Galaxy over the course of five years of
observations. We consider this model as an upper limit to the truly diffuse Galactic emission. In \S{sec:ani}, we include this
Galactic contribution and use both angular and spectral information to see how well the cosmological dark matter signal can be
extracted. We leave the 
spatial template and the spectral index fixed, and use only the normalization as a free parameter. 
This has a physical motivation: the spatial template---while realistically different from the above toy-model---will 
be obtained by high-statistics sub-GeV observations.
Since its shape depends on the product of density of interstellar material times cosmic-ray density along the line of sight,
one does not expect it to change with energy. Also, the spectral
index 2.7 is more or less what is observed in cosmic ray protons of 10-10000 GeV energy (which generate the photons in the 
energy range of interest), and photons produced  by $\pi_0$ via spallation follow the same power-law as the primaries.

In the next section, we explore the power of spectral discrimination, then add in angular discrimination, and finally 
turn to discrimination via distributions in a simple 2-component model. 

\section{Spectral Discrimination}
\label{sec:freq}

One way to extract the dark matter annihilation signal from astrophysical backgrounds is to exploit differences in the 
spectrum of each component. We first focus on the simple example where the shapes of the spectra are known and we fit 
the data for the two amplitudes. Generalizing to the more
realistic case of unknown shape parameters is straightforward, and we illustrate this at the end of this section by allowing the slope of the blazar spectrum 
and the mass of the dark matter particle to vary.
In this section, we neglect all angular information and treat both signal and background as isotropic on the sky. We break the gamma ray sky up into 
$N_e$ different energy bins (we will use $N_e=25$ bins logarithmically spaced in energy between 1 GeV and 300 GeV). For now, we assume that the 
likelihood of observing $(N_1,N_2,\ldots N_{N_e})$
photons in each of the energy bins is Gaussian:
\be
\mathcal{L}\propto \exp\left\{  -\frac{1}{2}\sum_{i=1}^{N_e}
\frac{\Big(N_i - N^s f^s_i -N^b f^b_i\Big)^2}{\sigma_i^2}
\right\},\eql{gauss}
\ee
where $N^s$ is the total number of expected counts due to the (dark matter) signal in all bins 
and $f^s_i$ the corresponding spectral
shape normalized so that $\sum_i f^s_i=1$, and $N^b$ and $f^b_i$ are the analogous quantities for
the background. The noise in the i$^{th}$ bin is $\sigma_i$.
To project the errors on the two free parameters in this model ($N^s$ and $N^b$), we compute the
curvature of the likelihood function, or the $2\times2$ Fisher matrix,
\bea
F_{\mu\nu} &\equiv& - \left\langle\frac{\partial^2 \ln\mathcal{L}}{\partial N^\mu
\partial N^\nu}\right\rangle
\vs
&=& \sum_{i=1}^{N_e} \frac{f^\mu_i f^\nu_i}{\sigma_i^2},
\eea
where $\mu,\nu$ run over signal and background.  
Consider the case where the noise is Poisson noise so that $\sigma_i^2=
N^b f^b_i+N^s f^s_i$. Then the Fischer matrix simplifies to
\be
F_{\mu\nu} = \sum_{i=1}^{N_e} \frac{f^\mu_i f^\nu_i}{N^b f^b_i + N^s f^s_i}.
\ee
The $F_{ss}$ component of this matrix is the inverse of the square of the 1-$\sigma$ projected error on the number of
signal events assuming the number of background events, $N^b$, is known in advance. This is called the {\it unmarginalized} error on $N_s$:
\be 
(\Delta N_s)_{{\rm unmarg}} = \left[ \sum_{i=1}^{N_e} \frac{f^s_i f^s_i}{N^b f^b_i + N^s f^s_i} \right]^{-1/2}
.\eql{unmarg}\ee
More relevant is the error when $N_b$ is allowed to vary freely. In that case, the marginalized error on $N_s$ is
$[(F^{-1})_{ss}]^{1/2}$. Explicitly,
\be
(\Delta N^s)_{\rm marg} = \frac{(\Delta N^s)_{\rm unmarg}}{\sqrt{1-r^2}}, 
\eql{das}\ee
where $r$ measures the extent to which the two spectra are orthogonal to one another:
\be 
r\equiv \frac{F_{sb}}{\sqrt{F_{ss}F_{bb}}}\eql{r}
.\ee

If the two spectra are very different, then $r$ is close to zero, and it is easy to extract the signal from the background. Quantitatively, in that limit, $(\Delta N_s)_{\rm marg} = (\Delta N_s)_{\rm unmarg}$. Notice from \ec{unmarg} that this error scales as $\sqrt{N^b}$ as naively expected (e.g., significance as defined in Ref.~\cite{Kuhlen:2007wv}), with the shape functions providing the precise numerical coefficient. If the spectra are similar, though, the marginalized error can become arbitrarily large as $r$ approaches one. 
Eq.~\Ec{das} offers a compact way to
assess how effectively a given experiment can separate signal from background using spectral information alone.

In the idealized case in which the spectral shape and normalization of the diffuse background from unresolved blazars are known in advance (from a detailed population study of resolved blazars, for example), we find that this technique can be used to determine the number of signal events from five years of observation by Fermi to an accuracy of $\Delta N^s =1270$. This is only 2\% tighter than the Poisson error $\Delta N_s = \sqrt{N_b} = 1289$. So if the background photons counts were known exactly, spectral information would add little discriminatory power. In the absence of such information, however, we are forced to marginalize over the normalization of the background. In that case, \ec{das} projects that the error goes up to $(\Delta N_s)_{\rm marg}=6277$. 
A simple way to interpolate between these two extremes -- marginalized and unmarginalized errors -- is to introduce a prior on the background number counts. This corresponds to multiplying the likelihood in \ec{gauss} by $\exp\left[{-(N_b-\bar N_B)^2/2\sigma_{N_b}^2}\right]$, or equivalently by adding $1/\sigma_{N_b}^2$ to the $bb$ component of the Fisher matrix.

\Sfig{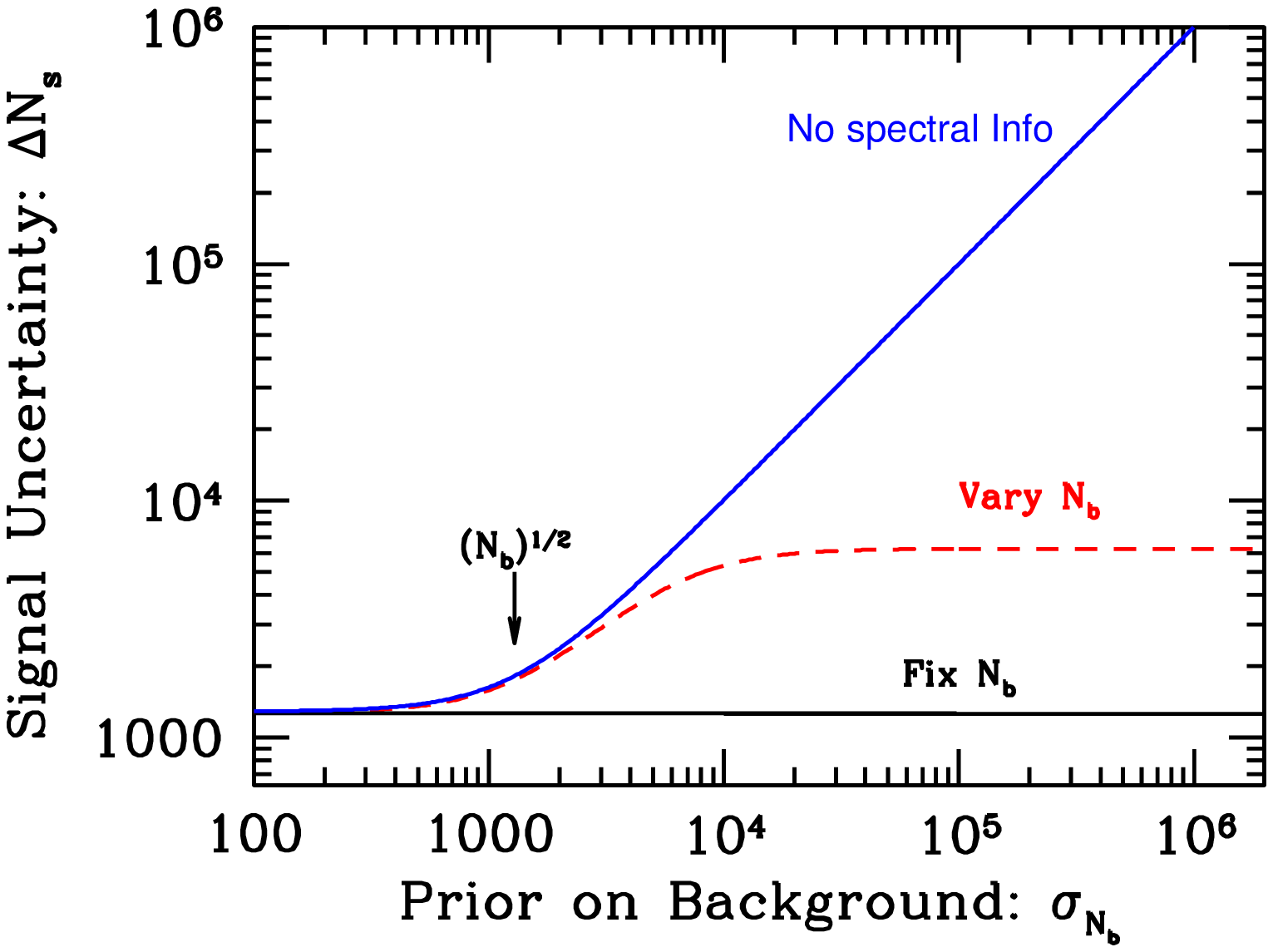}{The projected 1-sigma error on the number of events from dark matter annihilations as a function of how well known the background is for 5 years of Fermi observations. A Gaussian prior is placed on the number of background events with variance $\sigma_{N_b}^2$. The topmost line depicts the result if no spectral information is used; the middle line if spectral information from 25 bins is used; and the bottom horizontal line simply extends the ``fixed-background'' (corresponding to $\sigma_{N_b}=0$ result). Poisson noise -- the square root of the number of events -- is depicted by the vertical arrow. \label{signb}}

Fig.~\ref{signb} depicts the errors on $N_s$ as a function of the width of the prior, $\sigma_{N_b}$ ({\it ie.} 
the uncertainty on the background flux). If $\sigma_{N_b}$ is very small, much smaller than $N_b^{1/2}$, then the 
unmarginalized error is obtained. As the prior gets looser (larger $\sigma_{N_b}$), however, the projected error 
on $N_s$ gets larger. The middle (dashed) curve in Fig.~\ref{signb} illustrates the transition from the 
unmarginalized error to the marginalized result, about 4 times larger. The upper curve illustrates that, 
with no spectral discrimination, the error on $N_s$ scales simply as $\sigma_{N_b}$. The reality check here 
is that $N_b=1.7 \times 10^6$, so $\sigma_{N_b}\simeq1000$ -- roughly the transition region -- corresponds 
to knowing background counts to better than 0.1\%, clearly impossible. We thus conclude that, even with a 
very detailed blazar model derived from future population studies, we will not be able to predict the 
background flux with sufficient precision to make use of the unmarginalized error as described Eq.~\Ec{unmarg}. 
In all practical cases, analysts will need to marginalize over the background flux.

It is straightforward to vary other parameters, such as
the spectral index of the blazar spectrum (while still assuming a power law spectrum) and the mass of
the dark matter particle. The key ingredients in computing the Fisher matrix are the derivatives of the number of
events with respect to, now, the four parameters, taken
to be ln($N^s$), ln($N^b$), ln($m_{\rm DM}$), and $\alpha$, the slope of the
background spectrum. These derivatives are depicted in
Fig. \rf{md}.
  
\Sfig{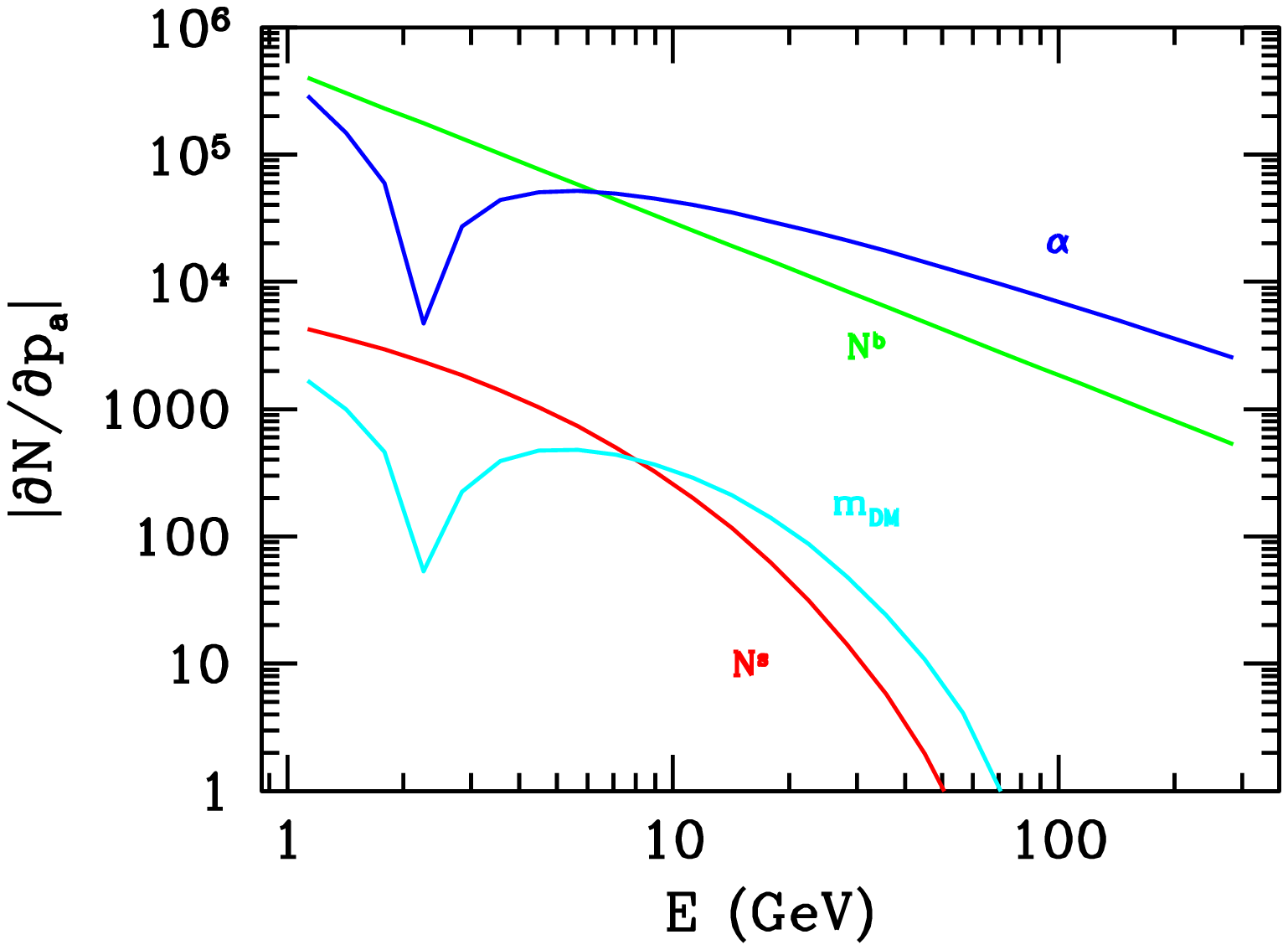}{The derivative of the total number of events in each
of 25 energy bins with respect to 4 parameters: ln($N^s$),
where $N^s$ is the number of photons from dark matter an-
nihilations; ln($N^b$), with $N^b$ the number of events from un-
resolved blazars; $\alpha$, the slope of the blazar spectrum; and
$m_{\rm DM}$, the dark matter mass. These derivatives are evaluated
around the fiducial values $(N^s,N^b,\alpha,m_{\rm DM}) = (1.9\times 10^4, 1.7 \times
10^6 , -2.2, 100\,{\rm GeV})$.\label{fig:md}}  

Marginalizing over the three other parameters
($N^b , \alpha, m_{\rm DM}$) leads to a 1-sigma error $\Delta N^s = 8846$ (as
opposed to 6277 found when the spectral index is fixed
to -2.2 and the mass to 100 GeV). Considering that an
NFW profile and a cross section of $\sigma v = 3 \times 10^{-26}$ cm$^3$
sec$^{-1}$ leads to 19,400 signal events, the 2-sigma upper limit
after 5 years would be $\simeq 2.7\times 10^{-26} {\rm cm}^3\, {\rm sec}^{-1}$ , consistent
with the results of Ref.~\cite{Baltz:2008wd}.

\section{Angular Discrimination}
\label{sec:ani}

Photons originating from cosmic rays incident on our Galaxy are likely to be far more numerous than those coming from outside the Galaxy. 
Indeed, in the model described in \S{\ref{sec:model}}, Fermi will detect $6.1\times 10^7$ Galactic photons over the course of 5 years over the whole sky.
This is almost 20 times larger than the number of photons produced by unresolved blazars and over a thousand times more than the extragalactic dark matter
signal\footnote{Recall that the numbers quoted in \S{\ref{sec:freq}} -- $1.7\times 10^6$ and 19,000 -- were for only half the sky. In this section we double these since we 
use the full sky.}. Spectral discrimination alone
will clearly not be sufficient to eliminate this background. Here we include the different angular distributions of the Galactic and extragalactic components
to project limits on the number of dark matter-produced events.

To include both angular and spectral information, we generalize the argument of the exponential in \ec{gauss} to 
\be
\chi^2 = \sum_{i=1}^{N_e} \sum_{a=1}^{N_{\rm pix}}
\frac{\Big(N_{i,a} - N^s f^s_i -N^b f^b_i - n^g I_{\rm gal}(E_i,l_a,b_a)\Big)^2}{\sigma_{ia}^2}
\eql{gaussani}
.\ee
Here, in addition to the sum over energy bins, we sum over $N_{\rm pix}$ angular pixels, each labeled with $(l_a,b_a)$. The model of \S\ref{sec:model} is multiplied by a normalization
factor $n^g$, equal to one in the model but allowed to float in our fit. The likelihood function (or $\chi^2$) therefore now depends on five parameters: two characterizing
the dark matter signal (amplitude $N^s$ and mass $m_{\rm DM}$); two characterizing extragalactic backgrounds (amplitude $N^b$ and slope $\alpha$); and one for the
normalization of the Galactic background $n^g$.

To project constraints on these parameters, we compute the (now 5-dimensional) Fisher matrix:
\be
F_{\mu\nu} = \frac{1}{2} \frac{\partial^2\chi^2}{\partial p^\mu\partial p^\nu}
\ee
where $p^\mu$ are the five parameters. For example, with $p^5=n^g$, taking the derivatives leads to 
\be
F_{55} = \sum_{i=1}^{N_e} \sum_{a=1}^{N_{\rm pix}}
\left(\frac{I_{\rm gal}(E_i,l_a,b_a)}{\sigma_{ia}}\right)^2.
\ee
The 1-sigma limit on the number of signal events, $\Delta N^s=\sqrt{(F^{-1})_{11}}$ is now equal to 34,000, very close to the full sky NFW signal of 39,000. The 2-sigma
upper limit on the annihilation cross section becomes $5.3\times 10^{-26}$ cm$^3$ sec$^{-1}$, so the Galactic photons pollute even regions far from the Galactic plane,
thereby degrading the upper limit by a factor of 2. 

\Sfig{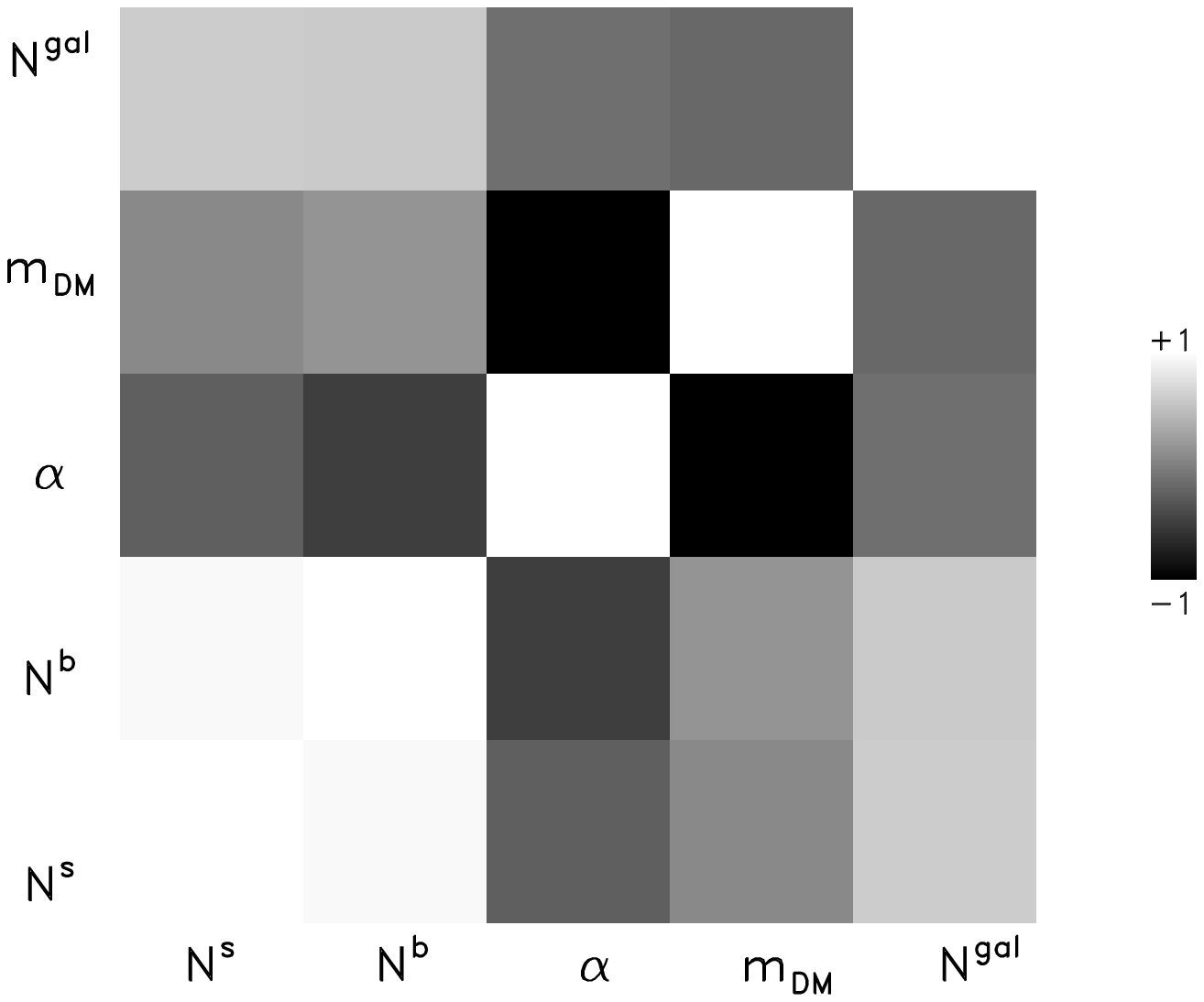}{The projected correlation matrix for a set of parameters used to fit 5 years of Fermi data. Note the strong correlation between $N^s$, the dark matter produced
photons, and the unresolved blazar background amplitude $N^b$. Similarly, the Galactic background is correlated
with $N^s$: $r_{15}=0.65$. Thus the Galactic photons degrade Fermi's sensitivity to this dark matter signal.\label{fig:rfish}}

The full Fisher matrix contains interesting information about the shape of the likelihood function in the full five dimensional parameter space. One way to explore this
structure is to generalize \ec{r} and consider the $5\times5$ dimensional correlation matrix with elements
\be
r_{\mu\nu} \equiv \frac{F_{\mu\nu}}{\sqrt{   F_{\mu\mu}  F_{\nu\nu}   }     }.
\ee
This is depicted in Fig.~\rf{rfish}. Note the strong correlation between the amplitudes of the isotropic components $N^s$ and $N^b$ and the 
strong anti-correlation between $m_{\rm DM}$ and $\alpha$ expected from the similarity in the derivatives in Fig.~\rf{md}.

\section{Distribution Discrimination}

As the distribution of photons from dark matter annihilations is expected to be close to Poisson, and the background from blazars is not, 
the natural question to ask is whether the signal can be  extracted from such backgrounds by exploiting this distinction. 
A complete answer to this question requires an understanding of the PDF's of all backgrounds and signals anf folding in
constraints from spectral and angular information such as those developed above. Here we take a first step in this direction by considering 
a toy model with just two components: extragalactic dark matter and unresolved blazars. Further we assume that the PDF of dark matter-produced photons is Poisson. 
As a preliminary illustration, note that with an average of 0.06 photons from dark matter annihilations in each $(0.25^{\circ})^2$ angular 
pixel, fewer than 0.5\% of all pixels will contain more than one photon from dark matter.  In contrast, 86\% (71\%) of all photons from blazars will fall in pixels with 10 (20) or more photons. Thus, by simply throwing away the photons in angular pixels with many photons, one can potentially remove the majority of the background from blazars, while retaining nearly all of the signal from dark matter.

Quantitatively, the probability of observing $\{N_1,N_2,\ldots\}$ 
photons in a set of $N_{\rm pix}$ pixels is given by
\be 
P\left[ \{ N_1, N_2, \ldots \} \vert N^s \right]
= \prod_{i=1}^{N_{\rm pix}} \sum_{j=0}^{N_i} P_b(N_i - j) P_s(j\vert N^s/N_{\rm pix}),
\eql{prob}\ee
where $P_b$ is the probability distribution for blazar photons, $P_s$ is the probability distribution for dark matter photons, 
and $N^s$ is the total number of signal photons expected (which scales with $\langle \sigma v \rangle$). 
$N^s$ is the only free parameter in the model. $P_s$ depends on the mean number of expected events 
in the pixel, equal to $N^s/N_{\rm pix}$. Here we do not use spectral information, so $N_1$ simply 
denotes the total number of photons detected in spatial pixel 1. The information contained in this distribution 
could be combined with spectral (and angular) information in a full likelihood analysis.

The standard assumption is to take both $P_b$ and $P_s$ to be Gaussian\footnote{This is virtually equivalent to taking the distributions to each be Poisson.}, so maximizing the likelihood reduces to minimizing the $\chi^2$:
\be 
\chi^2(N^s) \equiv \sum_{i=1}^{N_{\rm pix}} \frac{ \left( N_i - (N^s+N^b)/N_{\rm pix} \right)^2}{N_i},
\eql{chi2d}\ee
where $N^b$ is the total number of background photons and the denominator assumes that only Poisson noise is relevant. 
For the sake of this exercise, let us assume that $N^b$ is known. Under this 
assumption\footnote{When the uncertainty in $N^b$ is included, $\Delta N^s$ will go up as we saw in \S\ref{sec:freq}. The goal here though is to understand how much discrimination power lies in the different distributions, and we need a baseline prediction against which to judge the power, so we settle for fixed $N_b$.}, minimizing the $\chi^2$ leads to 
$\Delta N_s=\sqrt{N^b}$.

But what if the background counts were not drawn from a Gaussian distribution, but rather from the distribution shown in Fig.~\ref{distr}? 
How would this affect the results? Would an analyst who knew (or could estimate) the true distribution be able to exploit this information 
to extract the signal more effectively? Conversely, would an analyst ignorant of the true distribution who assumed a Gaussian distribution 
be led to false conclusions? To answer these questions, we generated counts in $N_{\rm pix}=330,000$ pixels (roughly ($0.25^\circ)^2$ each 
over half the sky) from the ``true'' distributions (Poisson for photons from dark matter, and that shown in Fig.~\ref{distr} for photons from blazars) 
and then analyzed these counts in two different ways in an attempt to extract the one free parameter, $N^s$. Then we repeated this exercise multiple 
times to accumulate statistics on how accurate each analysis technique was. The first technique analyzed the simulated data using the correct 
probability 
distributions in \ec{prob}, while the second assumed (incorrectly) that the backgrounds were also drawn from a Poisson distribution. In each case, 
we tabulated the likelihood function $\mathcal{L}(N^s) = P\left[ \{ N_1, N_2, \ldots \} \vert N^s \right]$ as a function of $N^s$ and computed the 
central 68\% confidence region. As expected, both analysis techniques retrieved the correct value of $N^s$ on average. The correct technique 
reported a 1-$\sigma$ error on $N^s$ of $331$; the Gaussian technique reported a 1-$\sigma$ error of 1291.  This is to be compared with the Poisson 
(unmarginalized) error of $\Delta N^s=1289$. We thus conclude that using the correct distribution leads to an improvement in sensitivity by a factor $\sim$4! 

\Sfig{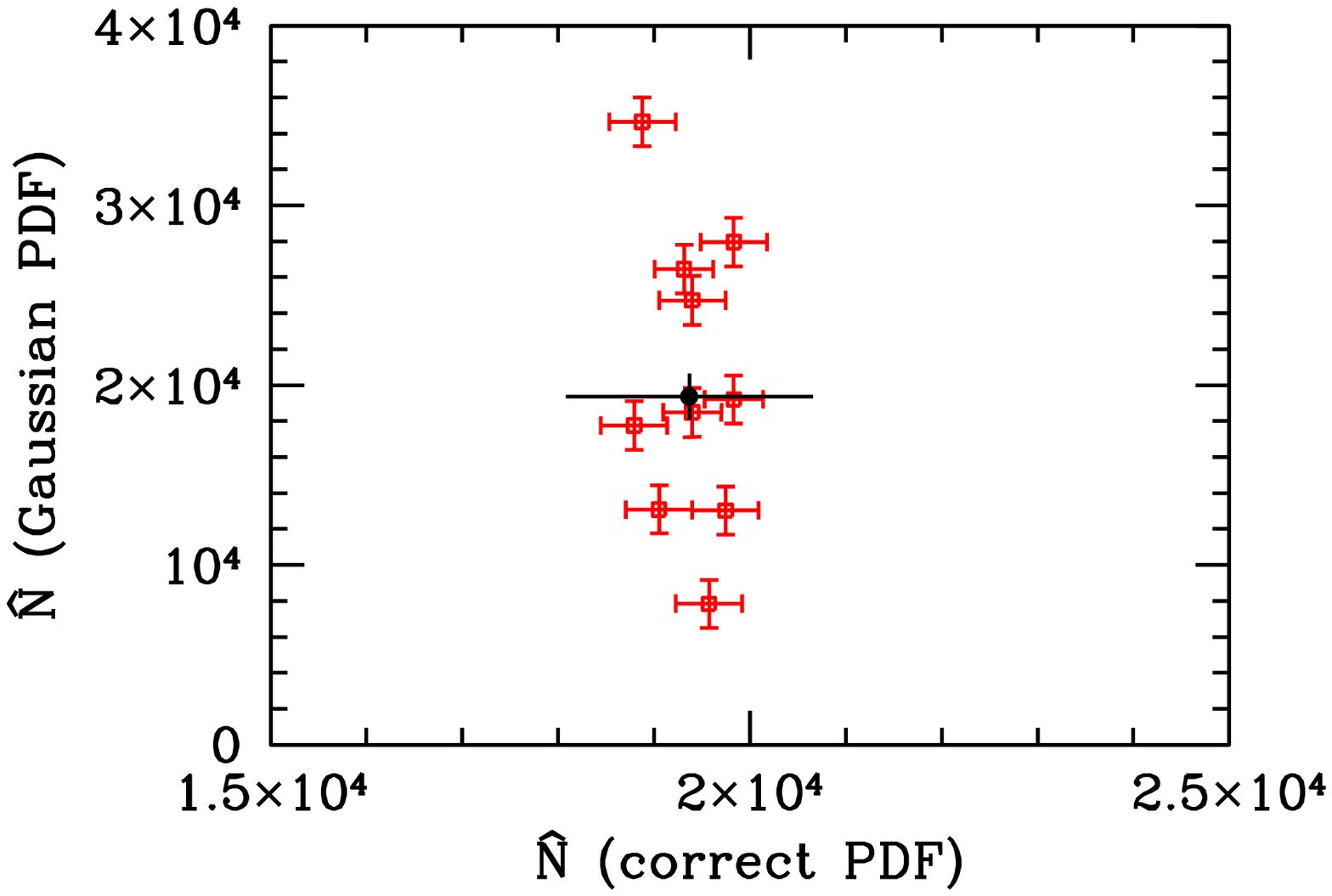}{Constraints on the number of events from dark matter annihilation from ten different simulations. Values along the x-axis were analyzed using the correct likelihood function in \ec{prob}, from which the simulations were drawn. Values along the y-axis were obtained by assuming (incorrectly) that the background events were drawn from a Gaussian distribution. Note the different scales along each axis. The black point is the true value and the error bars in each direction on that point represent Poisson errors in the background counts. Note that estimating $\hat N_s$ using the correct distribution leads to error bars smaller than Poisson and estimating it using the incorrect distribution leads to a large spread in the results.\label{sct}}

The corollary of the notion that knowing the underlying distributions is useful for extraction is the danger that {\it not knowing} the distributions 
will lead to errors. In fact, this happens when the incorrect distribution is assumed. Consider the results of the 10 runs depicted in Fig.~\ref{sct}. 
Each red box represents one Monte Carlo run analyzed with the two different likelihoods. The position of the box and the associated error bar along 
the horizontal axis denotes the estimate of $N^s$ and its 1-$\sigma$ error using the correct likelihood of \ec{prob}. The position of a box along 
the vertical axis, in contrast, denotes the estimate obtained using the (incorrect) Gaussian likelihood, similar to \ec{chi2d}. Note that the 
spread in the measurements using the correct estimator is comparable to the error bars. However, the spread in extracted values using the 
incorrect distribution is larger than the reported error bar by approximately an order of magnitude. This is a particularly
pernicious systematic error: if analysts unknowingly 
use the incorrect underlying distributions, the resulting estimates for $\Delta N^s$ will be much smaller than the true uncertainty. 
This result argues that, in order to optimally extract the dark matter signal, we need to understand the PDFs of both background and signal.

\section{Discussion and Conclusions}

In this article, we have studied the possibility of separating the cosmological gamma ray background produced in dark matter annihilations from the flux from unresolved blazars by using spectral information, angular information, and the differing probability distribution functions (PDFs). Using only spectral information, the resulting error on the
amplitude of the dark matter signal, given in \ec{das}, is a simple function of the spectra and the energy bins in the experiment. Angular information can/should also be incorporated to separate out the Galactic background. The probability distribution of pixel-counts from which the background and signal are drawn is also a potential discriminator. In particular, we have shown that the dark matter signal can be extracted
from a much larger background making use of these distributions. In fact, the extraction was even more effective than that obtained using spectral information, at least in the case considered here, providing a tool complementary to multipole analyses proposed in the recent past. The dangerous corollary of this result is that using an incorrect PDF can lead to a
systematic error in the signal extraction, potentially much larger than the corresponding statistical error. 

The analysis presented here has assumed two important approximations: (i) An isotropic background
resulting solely from unresolved blazars; and (ii) Photons from cosmological dark matter annihilations drawn from a Poisson distribution. 
It is currently believed that, at least well above one GeV, blazars are likely to be the main contributors to the unresolved gamma ray background 
(for a critical discussion of this point, see~\cite{Stecker:2001dk,Venters:2007rn,Dermer:2007fg}). Other backgrounds are also expected to be present including, 
for example, the ``guaranteed'' contribution from ordinary galaxies~\cite{Pavlidou:2002va} or the flux from byproducts of ultra-high energy cosmic ray 
interactions~\cite{Kalashev:2007sn} (for a review, see~\cite{Dermer:2007fg}).  
Depending on energy, these sources are expected to contribute from $\sim 0.1\%$ to $\sim 10\%$ of the EGRET background, and have a distribution closer 
to that from dark matter than from blazars. While the Galaxy contribution has a spectral shape quite different from the expected dark matter signal, 
the background from extragalactic cosmic ray interactions would be quite degenerate with it, making the method presented here unlikely to be 
successful in identifying the dark matter component if it is below a few percent of the EGRET diffuse flux. One might turn the argument around and 
conclude that, even in absence of a dark matter signal, the method presented here might be useful in studying sub-dominant, quasi-isotropic 
components of the diffuse signal. 
The second approximation mentioned above should prove easier to address. We can study the PDF of the dark matter signal as was done for 
Galactic sub-halos in Ref.~\cite{Lee:2008fm} to enhance the separation power. 
Furthermore, as population studies from Fermi become available, a more realistic model of unresolved blazars (as well as other potential 
gamma-ray sources) can be constructed.

As a final remark, let us stress that these considerations could significantly improve the bounds on {\it decaying} dark matter candidates as well. For a given particle physics scenario, the assumption of Poisson-distributed
cosmological emission should be an even better approximation; furthermore, the signal does not suffer from uncertainties of halo profiles and sub-structures. 
Further, in this case, the isotropic component is even more important for detection, since for decaying dark matter one does not expect a much larger signal from the Galactic Center region.
\bigskip

This work has been supported by the US Department of Energy, including grant DE-FG02-95ER40896 and by NASA grant
NAG5-10842. We thank Gianfranco Bertone, Savvas Koushiappas, and Louie Strigari for helpful comments.

\bibliography{indirect2}

\begin{thebibliography}{44}
\expandafter\ifx\csname natexlab\endcsname\relax\def\natexlab#1{#1}\fi
\expandafter\ifx\csname bibnamefont\endcsname\relax
  \def\bibnamefont#1{#1}\fi
\expandafter\ifx\csname bibfnamefont\endcsname\relax
  \def\bibfnamefont#1{#1}\fi
\expandafter\ifx\csname citenamefont\endcsname\relax
  \def\citenamefont#1{#1}\fi
\expandafter\ifx\csname url\endcsname\relax
  \def\url#1{\texttt{#1}}\fi
\expandafter\ifx\csname urlprefix\endcsname\relax\def\urlprefix{URL }\fi
\providecommand{\bibinfo}[2]{#2}
\providecommand{\eprint}[2][]{\url{#2}}

\bibitem[{\citenamefont{Bertone et~al.}(2005)\citenamefont{Bertone, Hooper, and
  Silk}}]{Bertone:2004pz}
\bibinfo{author}{\bibfnamefont{G.}~\bibnamefont{Bertone}},
  \bibinfo{author}{\bibfnamefont{D.}~\bibnamefont{Hooper}}, \bibnamefont{and}
  \bibinfo{author}{\bibfnamefont{J.}~\bibnamefont{Silk}},
  \bibinfo{journal}{Phys. Rept.} \textbf{\bibinfo{volume}{405}},
  \bibinfo{pages}{279} (\bibinfo{year}{2005}), \eprint{hep-ph/0404175}.

\bibitem[{\citenamefont{Aprile et~al.}(2005)}]{Aprile:2005ww}
\bibinfo{author}{\bibfnamefont{E.}~\bibnamefont{Aprile}} \bibnamefont{et~al.},
  \bibinfo{journal}{New Astron. Rev.} \textbf{\bibinfo{volume}{49}},
  \bibinfo{pages}{289} (\bibinfo{year}{2005}).

\bibitem[{\citenamefont{Sanglard}(2007)}]{Sanglard:2006hd}
\bibinfo{author}{\bibfnamefont{V.}~\bibnamefont{Sanglard}}
  (\bibinfo{collaboration}{EDELWEISS}), \bibinfo{journal}{Nucl. Phys. Proc.
  Suppl.} \textbf{\bibinfo{volume}{173}}, \bibinfo{pages}{99}
  (\bibinfo{year}{2007}), \eprint{astro-ph/0612207}.

\bibitem[{\citenamefont{Bisset et~al.}(2007)}]{Bisset:2007mz}
\bibinfo{author}{\bibfnamefont{R.}~\bibnamefont{Bisset}} \bibnamefont{et~al.},
  \bibinfo{journal}{Nucl. Phys. Proc. Suppl.} \textbf{\bibinfo{volume}{173}},
  \bibinfo{pages}{164} (\bibinfo{year}{2007}), \eprint{0705.2117}.

\bibitem[{\citenamefont{Bruch}(2007)}]{Bruch:2007zz}
\bibinfo{author}{\bibfnamefont{T.}~\bibnamefont{Bruch}}
  (\bibinfo{collaboration}{CDMS}), \bibinfo{journal}{AIP Conf. Proc.}
  \textbf{\bibinfo{volume}{957}}, \bibinfo{pages}{193} (\bibinfo{year}{2007}).

\bibitem[{\citenamefont{Baltz et~al.}(2006)\citenamefont{Baltz, Battaglia,
  Peskin, and Wizansky}}]{Baltz:2006fm}
\bibinfo{author}{\bibfnamefont{E.~A.} \bibnamefont{Baltz}},
  \bibinfo{author}{\bibfnamefont{M.}~\bibnamefont{Battaglia}},
  \bibinfo{author}{\bibfnamefont{M.~E.} \bibnamefont{Peskin}},
  \bibnamefont{and} \bibinfo{author}{\bibfnamefont{T.}~\bibnamefont{Wizansky}},
  \bibinfo{journal}{Phys. Rev.} \textbf{\bibinfo{volume}{D74}},
  \bibinfo{pages}{103521} (\bibinfo{year}{2006}), \eprint{hep-ph/0602187}.

\bibitem[{\citenamefont{Bergstrom et~al.}(1998)\citenamefont{Bergstrom, Ullio,
  and Buckley}}]{Bergstrom:1997fj}
\bibinfo{author}{\bibfnamefont{L.}~\bibnamefont{Bergstrom}},
  \bibinfo{author}{\bibfnamefont{P.}~\bibnamefont{Ullio}}, \bibnamefont{and}
  \bibinfo{author}{\bibfnamefont{J.~H.} \bibnamefont{Buckley}},
  \bibinfo{journal}{Astropart. Phys.} \textbf{\bibinfo{volume}{9}},
  \bibinfo{pages}{137} (\bibinfo{year}{1998}), \eprint{astro-ph/9712318}.

\bibitem[{\citenamefont{Cecchi}(2008)}]{Cecchi:2008zz}
\bibinfo{author}{\bibfnamefont{C.}~\bibnamefont{Cecchi}}
  (\bibinfo{collaboration}{GLAST LAT}), \bibinfo{journal}{J. Phys. Conf. Ser.}
  \textbf{\bibinfo{volume}{120}}, \bibinfo{pages}{062017}
  (\bibinfo{year}{2008}).

\bibitem[{\citenamefont{Dodelson et~al.}(2008)\citenamefont{Dodelson, Hooper,
  and Serpico}}]{Dodelson:2007gd}
\bibinfo{author}{\bibfnamefont{S.}~\bibnamefont{Dodelson}},
  \bibinfo{author}{\bibfnamefont{D.}~\bibnamefont{Hooper}}, \bibnamefont{and}
  \bibinfo{author}{\bibfnamefont{P.~D.} \bibnamefont{Serpico}},
  \bibinfo{journal}{Phys. Rev.} \textbf{\bibinfo{volume}{D77}},
  \bibinfo{pages}{063512} (\bibinfo{year}{2008}), \eprint{0711.4621}.

\bibitem[{\citenamefont{Miniati et~al.}(2007)\citenamefont{Miniati,
  Koushiappas, and Di~Matteo}}]{Miniati:2007ke}
\bibinfo{author}{\bibfnamefont{F.}~\bibnamefont{Miniati}},
  \bibinfo{author}{\bibfnamefont{S.~M.} \bibnamefont{Koushiappas}},
  \bibnamefont{and}
  \bibinfo{author}{\bibfnamefont{T.}~\bibnamefont{Di~Matteo}},
  \bibinfo{journal}{Astrophys. J.} \textbf{\bibinfo{volume}{667}},
  \bibinfo{pages}{L1} (\bibinfo{year}{2007}), \eprint{astro-ph/0702083}.

\bibitem[{\citenamefont{Hooper and Serpico}(2007)}]{Hooper:2007be}
\bibinfo{author}{\bibfnamefont{D.}~\bibnamefont{Hooper}} \bibnamefont{and}
  \bibinfo{author}{\bibfnamefont{P.~D.} \bibnamefont{Serpico}},
  \bibinfo{journal}{JCAP} \textbf{\bibinfo{volume}{0706}}, \bibinfo{pages}{013}
  (\bibinfo{year}{2007}), \eprint{astro-ph/0702328}.

\bibitem[{\citenamefont{Siegal-Gaskins}(2008)}]{SiegalGaskins:2008ge}
\bibinfo{author}{\bibfnamefont{J.~M.} \bibnamefont{Siegal-Gaskins}},
  \bibinfo{journal}{JCAP} \textbf{\bibinfo{volume}{0810}}, \bibinfo{pages}{040}
  (\bibinfo{year}{2008}), \eprint{0807.1328}.

\bibitem[{\citenamefont{Fornasa et~al.}(2009)\citenamefont{Fornasa, Pieri,
  Bertone, and Branchini}}]{Fornasa:2009qh}
\bibinfo{author}{\bibfnamefont{M.}~\bibnamefont{Fornasa}},
  \bibinfo{author}{\bibfnamefont{L.}~\bibnamefont{Pieri}},
  \bibinfo{author}{\bibfnamefont{G.}~\bibnamefont{Bertone}}, \bibnamefont{and}
  \bibinfo{author}{\bibfnamefont{E.}~\bibnamefont{Branchini}}
  (\bibinfo{year}{2009}), \eprint{0901.2921}.

\bibitem[{\citenamefont{Siegal-Gaskins and
  Pavlidou}(2009)}]{SiegalGaskins:2009ux}
\bibinfo{author}{\bibfnamefont{J.~M.} \bibnamefont{Siegal-Gaskins}}
  \bibnamefont{and} \bibinfo{author}{\bibfnamefont{V.}~\bibnamefont{Pavlidou}}
  (\bibinfo{year}{2009}), \eprint{0901.3776}.

\bibitem[{\citenamefont{Sreekumar et~al.}(1998)}]{Sreekumar:1997un}
\bibinfo{author}{\bibfnamefont{P.}~\bibnamefont{Sreekumar}}
  \bibnamefont{et~al.} (\bibinfo{collaboration}{EGRET}),
  \bibinfo{journal}{Astrophys. J.} \textbf{\bibinfo{volume}{494}},
  \bibinfo{pages}{523} (\bibinfo{year}{1998}), \eprint{astro-ph/9709257}.

\bibitem[{\citenamefont{Strong et~al.}(2004)\citenamefont{Strong, Moskalenko,
  and Reimer}}]{Strong:2004ry}
\bibinfo{author}{\bibfnamefont{A.~W.} \bibnamefont{Strong}},
  \bibinfo{author}{\bibfnamefont{I.~V.} \bibnamefont{Moskalenko}},
  \bibnamefont{and} \bibinfo{author}{\bibfnamefont{O.}~\bibnamefont{Reimer}},
  \bibinfo{journal}{Astrophys. J.} \textbf{\bibinfo{volume}{613}},
  \bibinfo{pages}{956} (\bibinfo{year}{2004}), \eprint{astro-ph/0405441}.

\bibitem[{\citenamefont{Elsaesser and Mannheim}(2005)}]{Elsaesser:2004ap}
\bibinfo{author}{\bibfnamefont{D.}~\bibnamefont{Elsaesser}} \bibnamefont{and}
  \bibinfo{author}{\bibfnamefont{K.}~\bibnamefont{Mannheim}},
  \bibinfo{journal}{Phys. Rev. Lett.} \textbf{\bibinfo{volume}{94}},
  \bibinfo{pages}{171302} (\bibinfo{year}{2005}), \eprint{astro-ph/0405235}.

\bibitem[{\citenamefont{de~Boer et~al.}(2005)\citenamefont{de~Boer, Sander,
  Zhukov, Gladyshev, and Kazakov}}]{deBoer:2006tv}
\bibinfo{author}{\bibfnamefont{W.}~\bibnamefont{de~Boer}},
  \bibinfo{author}{\bibfnamefont{C.}~\bibnamefont{Sander}},
  \bibinfo{author}{\bibfnamefont{V.}~\bibnamefont{Zhukov}},
  \bibinfo{author}{\bibfnamefont{A.~V.} \bibnamefont{Gladyshev}},
  \bibnamefont{and} \bibinfo{author}{\bibfnamefont{D.~I.}
  \bibnamefont{Kazakov}}, \bibinfo{journal}{Phys. Rev. Lett.}
  \textbf{\bibinfo{volume}{95}}, \bibinfo{pages}{209001}
  (\bibinfo{year}{2005}), \eprint{astro-ph/0602325}.

\bibitem[{\citenamefont{Ando}(2005)}]{Ando:2005hr}
\bibinfo{author}{\bibfnamefont{S.}~\bibnamefont{Ando}}, \bibinfo{journal}{Phys.
  Rev. Lett.} \textbf{\bibinfo{volume}{94}}, \bibinfo{pages}{171303}
  (\bibinfo{year}{2005}), \eprint{astro-ph/0503006}.

\bibitem[{\citenamefont{Ando and Komatsu}(2006)}]{Ando:2005xg}
\bibinfo{author}{\bibfnamefont{S.}~\bibnamefont{Ando}} \bibnamefont{and}
  \bibinfo{author}{\bibfnamefont{E.}~\bibnamefont{Komatsu}},
  \bibinfo{journal}{Phys. Rev.} \textbf{\bibinfo{volume}{D73}},
  \bibinfo{pages}{023521} (\bibinfo{year}{2006}), \eprint{astro-ph/0512217}.

\bibitem[{\citenamefont{Ando et~al.}(2007{\natexlab{a}})\citenamefont{Ando,
  Komatsu, Narumoto, and Totani}}]{Ando:2006mt}
\bibinfo{author}{\bibfnamefont{S.}~\bibnamefont{Ando}},
  \bibinfo{author}{\bibfnamefont{E.}~\bibnamefont{Komatsu}},
  \bibinfo{author}{\bibfnamefont{T.}~\bibnamefont{Narumoto}}, \bibnamefont{and}
  \bibinfo{author}{\bibfnamefont{T.}~\bibnamefont{Totani}},
  \bibinfo{journal}{Mon. Not. Roy. Astron. Soc.}
  \textbf{\bibinfo{volume}{376}}, \bibinfo{pages}{1635}
  (\bibinfo{year}{2007}{\natexlab{a}}), \eprint{astro-ph/0610155}.

\bibitem[{\citenamefont{Cuoco et~al.}(2007)}]{Cuoco:2006tr}
\bibinfo{author}{\bibfnamefont{A.}~\bibnamefont{Cuoco}} \bibnamefont{et~al.},
  \bibinfo{journal}{JCAP} \textbf{\bibinfo{volume}{0704}}, \bibinfo{pages}{013}
  (\bibinfo{year}{2007}), \eprint{astro-ph/0612559}.

\bibitem[{\citenamefont{Ando et~al.}(2007{\natexlab{b}})\citenamefont{Ando,
  Komatsu, Narumoto, and Totani}}]{Ando:2006cr}
\bibinfo{author}{\bibfnamefont{S.}~\bibnamefont{Ando}},
  \bibinfo{author}{\bibfnamefont{E.}~\bibnamefont{Komatsu}},
  \bibinfo{author}{\bibfnamefont{T.}~\bibnamefont{Narumoto}}, \bibnamefont{and}
  \bibinfo{author}{\bibfnamefont{T.}~\bibnamefont{Totani}},
  \bibinfo{journal}{Phys. Rev.} \textbf{\bibinfo{volume}{D75}},
  \bibinfo{pages}{063519} (\bibinfo{year}{2007}{\natexlab{b}}),
  \eprint{astro-ph/0612467}.

\bibitem[{\citenamefont{Cuoco et~al.}(2008)\citenamefont{Cuoco, Brandbyge,
  Hannestad, Haugboelle, and Miele}}]{Cuoco:2007sh}
\bibinfo{author}{\bibfnamefont{A.}~\bibnamefont{Cuoco}},
  \bibinfo{author}{\bibfnamefont{J.}~\bibnamefont{Brandbyge}},
  \bibinfo{author}{\bibfnamefont{S.}~\bibnamefont{Hannestad}},
  \bibinfo{author}{\bibfnamefont{T.}~\bibnamefont{Haugboelle}},
  \bibnamefont{and} \bibinfo{author}{\bibfnamefont{G.}~\bibnamefont{Miele}},
  \bibinfo{journal}{Phys. Rev.} \textbf{\bibinfo{volume}{D77}},
  \bibinfo{pages}{123518} (\bibinfo{year}{2008}), \eprint{0710.4136}.

\bibitem[{\citenamefont{Taoso et~al.}(2008)\citenamefont{Taoso, Ando, Bertone,
  and Profumo}}]{Taoso:2008qz}
\bibinfo{author}{\bibfnamefont{M.}~\bibnamefont{Taoso}},
  \bibinfo{author}{\bibfnamefont{S.}~\bibnamefont{Ando}},
  \bibinfo{author}{\bibfnamefont{G.}~\bibnamefont{Bertone}}, \bibnamefont{and}
  \bibinfo{author}{\bibfnamefont{S.}~\bibnamefont{Profumo}}
  (\bibinfo{year}{2008}), \eprint{0811.4493}.

\bibitem[{\citenamefont{Lee et~al.}(2008)\citenamefont{Lee, Ando, and
  Kamionkowski}}]{Lee:2008fm}
\bibinfo{author}{\bibfnamefont{S.~K.} \bibnamefont{Lee}},
  \bibinfo{author}{\bibfnamefont{S.}~\bibnamefont{Ando}}, \bibnamefont{and}
  \bibinfo{author}{\bibfnamefont{M.}~\bibnamefont{Kamionkowski}}
  (\bibinfo{year}{2008}), \eprint{0810.1284}.

\bibitem[{\citenamefont{Silk and Stebbins}(1993)}]{Silk:1992bh}
\bibinfo{author}{\bibfnamefont{J.}~\bibnamefont{Silk}} \bibnamefont{and}
  \bibinfo{author}{\bibfnamefont{A.}~\bibnamefont{Stebbins}},
  \bibinfo{journal}{Astrophys. J.} \textbf{\bibinfo{volume}{411}},
  \bibinfo{pages}{439} (\bibinfo{year}{1993}).

\bibitem[{\citenamefont{Ullio et~al.}(2002)\citenamefont{Ullio, Bergstrom,
  Edsjo, and Lacey}}]{Ullio:2002pj}
\bibinfo{author}{\bibfnamefont{P.}~\bibnamefont{Ullio}},
  \bibinfo{author}{\bibfnamefont{L.}~\bibnamefont{Bergstrom}},
  \bibinfo{author}{\bibfnamefont{J.}~\bibnamefont{Edsjo}}, \bibnamefont{and}
  \bibinfo{author}{\bibfnamefont{C.~G.} \bibnamefont{Lacey}},
  \bibinfo{journal}{Phys. Rev.} \textbf{\bibinfo{volume}{D66}},
  \bibinfo{pages}{123502} (\bibinfo{year}{2002}), \eprint{astro-ph/0207125}.

\bibitem[{\citenamefont{Navarro et~al.}(1997)\citenamefont{Navarro, Frenk, and
  White}}]{Navarro:1996gj}
\bibinfo{author}{\bibfnamefont{J.~F.} \bibnamefont{Navarro}},
  \bibinfo{author}{\bibfnamefont{C.~S.} \bibnamefont{Frenk}}, \bibnamefont{and}
  \bibinfo{author}{\bibfnamefont{S.~D.~M.} \bibnamefont{White}},
  \bibinfo{journal}{Astrophys. J.} \textbf{\bibinfo{volume}{490}},
  \bibinfo{pages}{493} (\bibinfo{year}{1997}), \eprint{astro-ph/9611107}.

\bibitem[{\citenamefont{Moore et~al.}(1999)}]{Moore:1999nt}
\bibinfo{author}{\bibfnamefont{B.}~\bibnamefont{Moore}} \bibnamefont{et~al.},
  \bibinfo{journal}{Astrophys. J.} \textbf{\bibinfo{volume}{524}},
  \bibinfo{pages}{L19} (\bibinfo{year}{1999}).

\bibitem[{\citenamefont{Bullock et~al.}(2001)}]{Bullock:1999he}
\bibinfo{author}{\bibfnamefont{J.~S.} \bibnamefont{Bullock}}
  \bibnamefont{et~al.}, \bibinfo{journal}{Mon. Not. Roy. Astron. Soc.}
  \textbf{\bibinfo{volume}{321}}, \bibinfo{pages}{559} (\bibinfo{year}{2001}),
  \eprint{astro-ph/9908159}.

\bibitem[{\citenamefont{Hartman et~al.}(1999)}]{Hartman:1999fc}
\bibinfo{author}{\bibfnamefont{R.~C.} \bibnamefont{Hartman}}
  \bibnamefont{et~al.} (\bibinfo{collaboration}{EGRET}),
  \bibinfo{journal}{Astrophys. J. Suppl.} \textbf{\bibinfo{volume}{123}},
  \bibinfo{pages}{79} (\bibinfo{year}{1999}).

\bibitem[{\citenamefont{Mukherjee et~al.}(1997)}]{Mukherjee:1997qw}
\bibinfo{author}{\bibfnamefont{R.}~\bibnamefont{Mukherjee}}
  \bibnamefont{et~al.}, \bibinfo{journal}{Astrophys. J.}
  \textbf{\bibinfo{volume}{490}}, \bibinfo{pages}{116} (\bibinfo{year}{1997}).

\bibitem[{\citenamefont{Chiang and Mukherjee}(1998)}]{luminosityfunction}
\bibinfo{author}{\bibfnamefont{J.}~\bibnamefont{Chiang}} \bibnamefont{and}
  \bibinfo{author}{\bibfnamefont{R.}~\bibnamefont{Mukherjee}},
  \bibinfo{journal}{Astrophys. J.} \textbf{\bibinfo{volume}{496}},
  \bibinfo{pages}{752} (\bibinfo{year}{1998}).

\bibitem[{\citenamefont{Dermer}(2007{\natexlab{a}})}]{Dermer:2006pd}
\bibinfo{author}{\bibfnamefont{C.~D.} \bibnamefont{Dermer}},
  \bibinfo{journal}{Astrophys. J.} \textbf{\bibinfo{volume}{659}},
  \bibinfo{pages}{958} (\bibinfo{year}{2007}{\natexlab{a}}),
  \eprint{astro-ph/0605402}.

\bibitem[{\citenamefont{Abdo et~al.}(2009)}]{Abdo:2009wu}
\bibinfo{author}{\bibfnamefont{A.~A.} \bibnamefont{Abdo}} \bibnamefont{et~al.}
  (\bibinfo{collaboration}{Fermi LAT}) (\bibinfo{year}{2009}),
  \eprint{0902.1559}.

\bibitem[{\citenamefont{Hunter et~al.}(1997)}]{Hunter:1997}
\bibinfo{author}{\bibfnamefont{S.~D.} \bibnamefont{Hunter}}
  \bibnamefont{et~al.}, \bibinfo{journal}{Astrophys. J.}
  \textbf{\bibinfo{volume}{481}}, \bibinfo{pages}{205} (\bibinfo{year}{1997}).

\bibitem[{\citenamefont{Kuhlen et~al.}(2007)\citenamefont{Kuhlen, Diemand, and
  Madau}}]{Kuhlen:2007wv}
\bibinfo{author}{\bibfnamefont{M.}~\bibnamefont{Kuhlen}},
  \bibinfo{author}{\bibfnamefont{J.}~\bibnamefont{Diemand}}, \bibnamefont{and}
  \bibinfo{author}{\bibfnamefont{P.}~\bibnamefont{Madau}},
  \bibinfo{journal}{AIP Conf. Proc.} \textbf{\bibinfo{volume}{921}},
  \bibinfo{pages}{135} (\bibinfo{year}{2007}), \eprint{0704.0944}.

\bibitem[{\citenamefont{Baltz et~al.}(2008)}]{Baltz:2008wd}
\bibinfo{author}{\bibfnamefont{E.~A.} \bibnamefont{Baltz}}
  \bibnamefont{et~al.}, \bibinfo{journal}{JCAP}
  \textbf{\bibinfo{volume}{0807}}, \bibinfo{pages}{013} (\bibinfo{year}{2008}),
  \eprint{0806.2911}.

\bibitem[{\citenamefont{Stecker and Salamon}(2001)}]{Stecker:2001dk}
\bibinfo{author}{\bibfnamefont{F.~W.} \bibnamefont{Stecker}} \bibnamefont{and}
  \bibinfo{author}{\bibfnamefont{M.~H.} \bibnamefont{Salamon}}
  (\bibinfo{year}{2001}), \eprint{astro-ph/0104368}.

\bibitem[{\citenamefont{Venters and Pavlidou}(2007)}]{Venters:2007rn}
\bibinfo{author}{\bibfnamefont{T.~M.} \bibnamefont{Venters}} \bibnamefont{and}
  \bibinfo{author}{\bibfnamefont{V.}~\bibnamefont{Pavlidou}},
  \bibinfo{journal}{AIP Conf. Proc.} \textbf{\bibinfo{volume}{921}},
  \bibinfo{pages}{163} (\bibinfo{year}{2007}), \eprint{0704.2417}.

\bibitem[{\citenamefont{Dermer}(2007{\natexlab{b}})}]{Dermer:2007fg}
\bibinfo{author}{\bibfnamefont{C.~D.} \bibnamefont{Dermer}},
  \bibinfo{journal}{AIP Conf. Proc.} \textbf{\bibinfo{volume}{921}},
  \bibinfo{pages}{122} (\bibinfo{year}{2007}{\natexlab{b}}),
  \eprint{0704.2888}.

\bibitem[{\citenamefont{Pavlidou and Fields}(2002)}]{Pavlidou:2002va}
\bibinfo{author}{\bibfnamefont{V.}~\bibnamefont{Pavlidou}} \bibnamefont{and}
  \bibinfo{author}{\bibfnamefont{B.~D.} \bibnamefont{Fields}},
  \bibinfo{journal}{Astrophys. J.} \textbf{\bibinfo{volume}{575}},
  \bibinfo{pages}{L5} (\bibinfo{year}{2002}), \eprint{astro-ph/0207253}.

\bibitem[{\citenamefont{Kalashev et~al.}(2007)\citenamefont{Kalashev, Semikoz,
  and Sigl}}]{Kalashev:2007sn}
\bibinfo{author}{\bibfnamefont{O.~E.} \bibnamefont{Kalashev}},
  \bibinfo{author}{\bibfnamefont{D.~V.} \bibnamefont{Semikoz}},
  \bibnamefont{and} \bibinfo{author}{\bibfnamefont{G.}~\bibnamefont{Sigl}}
  (\bibinfo{year}{2007}), \eprint{0704.2463}.

\end{thebibliography}
\end{document}